\pdfoutput=1
\documentclass{article}[]
\usepackage[utf8]{inputenc}
\usepackage{graphicx}
\usepackage{url}
\usepackage{float}
\usepackage{authblk}

\graphicspath{ {./Images/} }

\title{Sewage Pooling Test for SARS-CoV-2}
   
\author[1]{Ritam Guha}
\author[2]{Anik Sengupta\textsuperscript{*}}
\author[3]{Ankan Dutta\thanks{authors contributed equally}}

\affil[1]{\small ritamguha16@ieee.org, Department of Computer Science and Engineering, Jadavpur University, India}
\affil[2]{\small senguptaanik8@gmail.com, Department of Electronics and Communication Engineering, Heritage Institute of Technology, India} 
\affil[3]{\small ankan19d@gmail.com, Department of Mechanical Engineering, Jadavpur University, India}

\date{May 2020} 
\begin{document}
\maketitle

\begin{abstract}
CoVID-19 is currently one of the biggest threats to mankind. To date, it is the reason for infections of over 35 lakhs and death of over 2 lakh human beings. We propose a procedure to detect CoVID-19 affected localities using a sewage mass testing and pooling mechanism which has gained ground in recent times. The proposed method named Sewage Pooling Algorithm tests wastewater samples from sewage systems to pinpoint the regions which are affected by maximum chances of the virus spread. The algorithm also uses a priority-based backtracking procedure to perform testing in sewage links depending on the probability of infection in the sub-areas. For places with very rare CoVID cases, we present a gradient-based search method to prune those areas. The proposed method has less human intervention and increases the effective tests/million people over current in-place methods.
\end{abstract}


\newpage

\section{Introduction}

The main routes of transmission of CoVID-19 are respiratory droplets and direct  contact. Current evidence  suggests  that the virus  may  be  excreted   in   faeces,   regardless   of   diarrhoea   or   signs of   intestinal  infection. Studies have shown that CoVID-19 viral RNA fragments can be detected more accurately from patient's (even asymptomatic patient's) stool specimen \cite{tang2020detection}\\ 

No one knows about the actual number of people affected with COVID-19. We are only aware of the status of those infectious people who have been tested. Testing is our only window to handle this pandemic situation as we get aware of the way virus is spreading and also actual number of infectious people.In the asymptomatic infectious cases in the community or when people are not sure whether they are infected or not,  community sewage detection could determine whether there are CoVID-19 carriers in an area so that effective intervention can be taken as soon as possible. The present testing mechanism is too slow, cost prohibitive and sometimes provide negative results even if the person is infectious. As of now in INDIA, number of samples being tested is 0.03 per thousand people \cite{roser2020coronavirus}. The proposed mechanism of wide sewage testing is daunting yet more manageable. Studies have shown that SARS COV-2 is highly similar to SARS COV-1 genetically \cite{wang2020genetic}. Now because of the genetic similarity, detection method remains quite similar, and there have been studies regarding  the detection method of SARS COV-1 from sewage samples \cite{wang2005concentration}.

\section{Proposed Model}

\subsection{Sewage System Structure}
\begin{figure}[!h]
    \centering
    \resizebox{\textwidth}{!}{
    \includegraphics[scale=0.5]{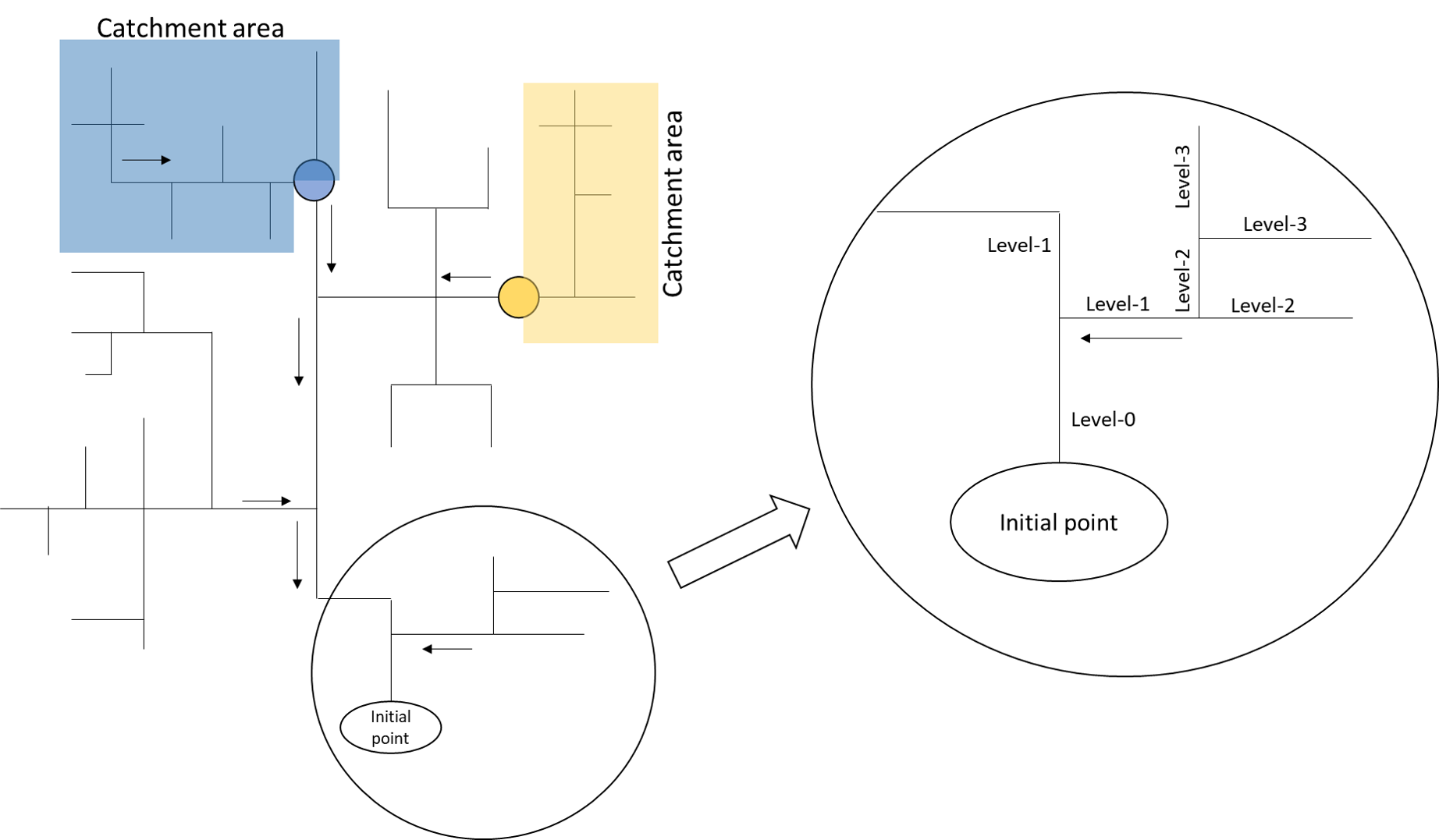}
    }
    \caption{Schematic diagram representing a Sewage System}    \label{fig:sewage system}
\end{figure}

To understand the proposed methodology, it is essential to get an overview of the sewage structure adapted in various places. The most common schematic diagram of a sewage system is presented in Figure \ref{fig:sewage system}. From the structure, it can be seen that many links are interconnected to form a backbone of the sewage system.These links carry waster water from different areas to the backbone structure which is finally dumped in a place which is marked as Initial Point in the Figure. Sometimes, this initial point is a Waster Water Treatment Plant (WWTP) which refines the waste water and passes it to rivers, ponds etc. 

At any point of any link, the entire area whose wastewater passes through that point is known as the catchment area of the point. So, a particular link is responsible to carry the waste water of its catchment area. For easy explanation and references, the proposed method labels the links in terms of a labelling system. The links emanating from the initial point are $level-0$ links which further leads to the subsequent levels of links ($level-1, level-2,\ldots, level-n$). All the $level-(i+1)$ links connected to a $level-i$ link are called the child links of that $level-i$ while $level-i$ link is called the parent link of its child links.

\subsection{Sewage Collection \& Testing}

In the present scenario, composite or integrated sewage samples are collected with sampling covering a period usually no longer than $24$ hrs using an automatic or mechanical sampler. Throughout this report we assume the sample taken at any link to be an equally probable sewage collection of any individual of the sample population of that region represented as the catchment area of that link. To facilitate this statement with higher certainty, we need to assure proper mixing of sewage via making the flow more turbulent and by increasing the sampling frequency and volume. Satisfactory results are also obtained by collecting a portion of uniform size each time a predetermined amount of flow had passed the sampling station \cite{Sampling1961}.\\

For detection of CoVID-19 from sewage, rapid test kit cannot be used as most of them are based on antibody test. Through the faeces of the human body, RNA fragments of the virus are excreted and not the antibody generated inside the body. One-step quantitative reverse transcription PCR (RT-qPCR) is decided to be used to achieve a faster and highly reproducible test. Recent studies of sewage test using RT-qPCR has shown better prediction of CoVID cases and is currently being used in few countries \cite{Wu2020, biobot2020, cnn2020}. One such is approved by ICMR is TRUPCR® SARS-CoV-2 RT-qPCR kit, based on real-time PCR technology, for the qualitative detection of (SARS-CoV-2) specific RNA \cite{trupcr2020}. \\

The precision of RT-qPCR is important to determine the largest sampling population the method can encounter or in other words, the initial point at which the pooling method in sewage context can start iterating. Suppose a sewage sample of a CoVID positive person is tested with $t$ viral particles per ml. We add the sample with sewage of other $n-1$ non-CoVID patients and test the diluted sample in RT-qPCR which will roughly show $t/n$ viral particles per ml. This value decreases with increasing the sample size $n$ till we get a maximal $n^*$ till which the equipment can detect the virus. We then group the region with a maximum pool size $n^*$. The initial points will be links in sewage system which can capture the population of $n^*$ independently. 

\subsection{Database updation}
For proper functioning of the proposed procedure, it is important to retrieve the latest set of information available in the market. Some very important parameters like number of Covid patients already present in the locality, rate of increase in the catchment area, number of infants and aged inhabitants of the area, population density of area etc. may play a very important role in selecting the most optimal set of paths for testing. So, the process needs access to the latest information in terms of these parameters and current test results. This should be done by a system which updates the information after regular intervals.\\   

\subsection{Sewage Pooling Algorithm}
The Covid-19 pandemic requires every patient to recover from the disease without affecting others around. Even the presence of a single patient may result into exponential spread of the virus thereby affecting a large population. So, the main aim of the proposed approach is to make an attempt to identify every Covid-19 affected locality. For this reason, the proposed model tries to collect test samples from every sewage link possible and test them but the selection of the links is not done blindly. It uses a probability-guided backtracking approach to find the most probable links in the current situation.\\

\begin{figure}[H]
    \centering
    \includegraphics[scale=0.5]{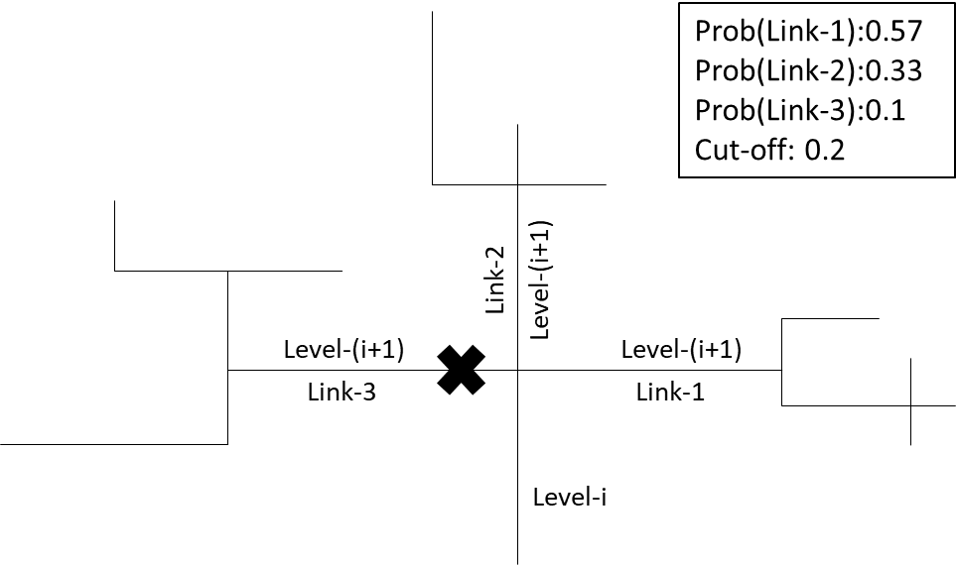}
    \caption{Example of the testing procedure}   
    \label{fig:pruning}
\end{figure}

After finding the starting point of the search procedure, the probability is calculated for each and every link in the area based on the information extracted from the demo-graphical study of the locality. Then the testing starts with the $level-0$ links. At any stage, if the probability is less than a certain cutoff, that particular branch is not tested (pruned) for the time being. Otherwise, the branches are tested in decreasing order of their probabilities. The procedure is trivially described in Figure \ref{fig:pruning}. $Link-1$ is tested first because it has the maximum probability of infection followed by $Link-2$. $Link-3$ is pruned because its probability value is below the cut-off.

\begin{figure}[!h]
    \centering
      \includegraphics[scale=0.5]{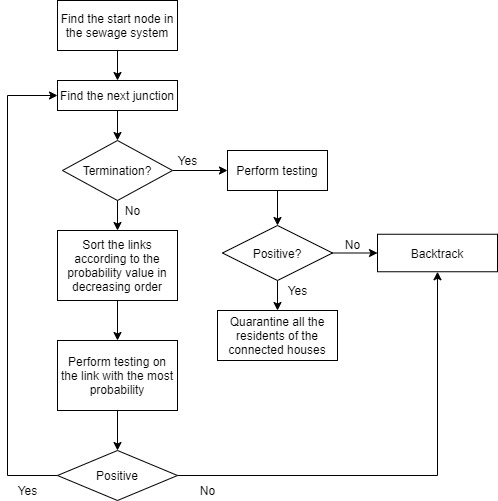}
    \caption{Flowchart of Priority-based backtracking procedure}    
    \label{fig:flochart_PB}
\end{figure}

\begin{figure}[!h]
    \centering
      \includegraphics[scale=0.5]{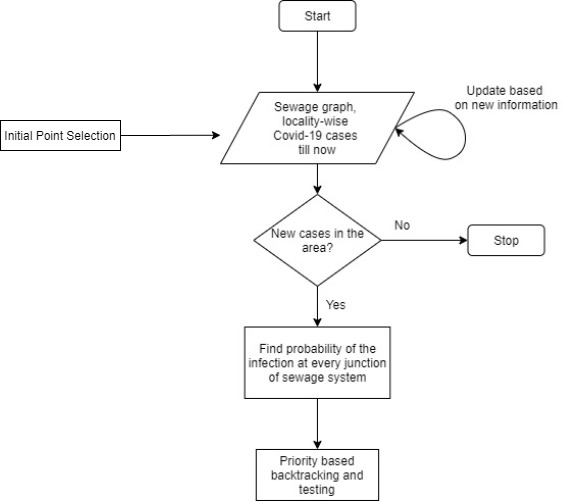}
    \caption{Flowchart of SPA}    
    \label{fig:flochart_SPA}
\end{figure}

When $level-(i+1)$ links are done with the testing, subsequent levels are tested using the same rule until terminal links are tested in a recursive manner. This procedure is known as Priority-based backtracking as probability is guiding this searching procedure.The flowchart for the entire procedure is provided in Figure \ref{fig:flochart_PB}. The entire approach is named as Sewage Pooling Algorithm presented in Figure \ref{fig:flochart_SPA}.    

Regarding quarantining, there are mainly two scenarios:
\begin{enumerate}
    \item If the search procedure ends at a terminal link which is tested positive, people in its catchment area should be quarantined and confirmatory tested.
    \item If an intermediate node is tested positive, then all the people staying at the unique houses only connected to that intermediate node must be tested.
\end{enumerate}

\paragraph{Gradient Searching:}
The choice of initial point is dependent on the sensitivity of RT-qPCR. In case of very low concentration or probability area which are being pruned, there will still be cases of CoVID which cannot be ignored in the long run. Moreover there may be exception when the sample doesn't really abstract the sewage information of the catchment area with equal probability. To tackle these problems we use gradient i.e. difference in concentration across depth-wise links. Suppose to test a sample at link $level-{n^*}$ or $level-0$ which is abstracting the sewage content of $n^*$ population with $m$ CoVID positive cases where $m<<n^*$ and the mixing doesn't equilibrate thus over-estimating $n^*$. We find gradient of all links of $level-1$ if the concentration at $level-0$ is less than a threshold (very low) above which the region is denoted as yellow zone in sewage testing context. As the CoVID cases are less, if the gradient is negative and the concentration value is less than the precision value, there is no CoVID case in that catchment region. This is because the sample size at $level - 1$ is less than $level - 0$ or $n^*$. So if there was a positive CoVID case there must be a rise in the CoVID level as a percentage. Moreover the concentration must not be less than precision if even one CoVID case is there as the sample size is much less than $n^*$. We prune the links where gradient is negative and move forward with positive gradient showing the direction.

\section{Unique Features}
This approach has many advantages over the currently available methods. The advantages are outlined in the following portion.
\begin{enumerate}
    \item Currently the process-in-action for CoVID treatment requires affected people to identify the symptoms, consult doctors, visit hospitals and then if positively identified, go for treatment> During this entire process, some more people may get affected if the person is positive. But, the proposed approach should be performed at Govt. level with \textbf{little to no effort from the affected} patients. 
    \item This process can \textbf{detect asymptomatic patients} as well.
    \item \textbf{Reduced number of test} are required on an average basis. (There may be scenarios in which most of the patients are affected in the locality. In this exceptional scenario, number of tests will increase)
    \item Even after the pandemic is over, this process can be used to keep a \textbf{track of CoVID} in different localities.
\end{enumerate}

\section{Conclusion}
This work presents an approach to identify the CoVID affected localities using a sewage pooling mechanism. The proposed method known as Sewage Pooling Algorithm tries to identify the affected areas by performing RT-qPCR testing on sewage samples collected from these areas in an informative way. It uses the data extracted from demo-graphic information of the regions to compute the probability of infections in different sub-areas connected by sewage links. Priority-based backtracking approach is used to perform testing in sub-areas with higher probability of infection. Although, to the best of our knowledge, it is a very efficient way to detect affected areas using less number of tests, it assumes ideal scenarios in many situations. For example, it requires proper mixing of the pooled sewage samples but in reality it may not be the case. There may be various organizations in the locality which negatively affects the results. For example, if the area has a hospital treating CoVID patients or a factory emanating chemicals which may change the nature of the wastewater samples, the result will get badly affected. The procedure needs some modifications to make it applicable to real-life situations which is not possible without practical testing and implementations. In future, if this idea gets approved, modifications can be performed over the procedure based on the new set of information obtained from the practical testing outcomes.  

\section{Conflict of Interest}
The authors have conflict of interest to declare. AS, AD presented this idea as QuickSolutions to Techstar India start-up weekend and secured the first prize in India Chapter.

\bibliographystyle{plain}
\bibliography{references.bib}

\begin{thebibliography}{1}

\bibitem{trupcr2020}
Trupcr® corona virus/covid-19 (sars-cov-2) real time pcr kits.
\newblock \url{https://www.3bblackbio.com/covid-19.html}.
\newblock Accessed: 2010-05-02.

\bibitem{cnn2020}
CNN.
\newblock Some scientists are using sewage to measure the prevalence of
  coronavirus in their communities.
\newblock
  \url{https://edition.cnn.com/2020/04/26/us/covid-19-sewage-testing/index.html}.
\newblock Accessed: 2010-05-02.

\bibitem{Sampling1961}
Water~Environment Federation.
\newblock Sampling of sewage.
\newblock {\em Journal (Water Pollution Control Federation)}, 33(6):669--674,
  1961.

\bibitem{biobot2020}
MIT.
\newblock Covid-19 testing in city sewage.
\newblock \url{https://www.biobot.io/covid19}.
\newblock Accessed: 2010-05-02.

\bibitem{roser2020coronavirus}
Max Roser, Hannah Ritchie, Esteban Ortiz-Ospina, and Joe Hasell.
\newblock Coronavirus disease (covid-19).
\newblock {\em Our World in data}, 2020.

\bibitem{tang2020detection}
An~Tang, ZD~Tong, HL~Wang, YX~Dai, KF~Li, JN~Liu, WJ~Wu, C~Yuan, ML~Yu, P~Li,
  et~al.
\newblock Detection of novel coronavirus by rt-pcr in stool specimen from
  asymptomatic child, china.
\newblock {\em Emerging infectious diseases}, 26(6), 2020.

\bibitem{wang2020genetic}
Huihui Wang, Xuemei Li, Tao Li, Shubing Zhang, Lianzi Wang, Xian Wu, and
  Jiaqing Liu.
\newblock The genetic sequence, origin, and diagnosis of sars-cov-2.
\newblock {\em European Journal of Clinical Microbiology \& Infectious
  Diseases}, page~1, 2020.

\bibitem{wang2005concentration}
Xin-Wei Wang, Jin-Song Li, Ting-Kai Guo, Bei Zhen, Qing-Xin Kong, Bin Yi, Zhong
  Li, Nong Song, Min Jin, Wen-Jun Xiao, et~al.
\newblock Concentration and detection of sars coronavirus in sewage from xiao
  tang shan hospital and the 309th hospital.
\newblock {\em Journal of virological methods}, 128(1-2):156--161, 2005.

\bibitem{Wu2020}
Fuqing Wu, Amy Xiao, Jianbo Zhang, Xiaoqiong Gu, Wei~Lin Lee, Kathryn Kauffman,
  William Hanage, Mariana Matus, Newsha Ghaeli, Noriko Endo, Claire Duvallet,
  Katya Moniz, Timothy Erickson, Peter Chai, Janelle Thompson, and Eric Alm.
\newblock {SARS}-{CoV}-2 titers in wastewater are higher than expected from
  clinically confirmed cases.
\newblock {\em medrXiv}, April 2020.

\end{thebibliography}

\end{document}